\documentclass{jps-cp}
\usepackage{txfonts} %Please comment out this line unless the txfonts package is availabe in your LaTeX system.

\title{Monte Carlo Simulations for the Optimization and Data Analysis of Experiments with Ultracold Neutrons}

\author{Nicholas J. \textsc{Ayres}$^{1}$, Estelle \textsc{Chanel}$^{1}$, Benoit \textsc{Clement}$^{2}$, Philip G. \textsc{Harris}$^{1}$, R\"udiger \textsc{Picker}$^{3}$, Guillaume \textsc{Pignol}$^{2}$, Wolfgang \textsc{Schreyer}$^{3}$ and Geza \textsc{Zsigmond}$^{4}$}

\inst{
$^{1}$Department of Physics and Astronomy, University of Sussex, Falmer, Brighton BN1 9QH, United Kingdom \\
$^{2}$LPSC Grenoble, Université Grenoble-Alpes, CNRS/IN2P3, 38026 Grenoble Cedex, France \\
$^{3}$TRIUMF, Physical Sciences Division, Vancouver, BC V6T 2A3, Canada  \\
$^{4}$Paul Scherrer Institut, CH-5232 Villigen PSI, Switzerland
}

\email{geza.zsigmond@psi.ch}

\recdate{October 2, 2017}

\abst{Ultracold neutrons (UCN) with kinetic energies up to 300 neV can be stored in material or magnetic confinements for hundreds of seconds. This makes them a very useful tool for probing fundamental symmetries of nature, by searching for charge-parity violation by a neutron electric dipole moment, and yielding important parameters for Big Bang nucleosynthesis, e.g. in neutron-lifetime measurements. Further increasing the intensity of UCN sources is crucial for next-generation experiments. Advanced Monte Carlo (MC) simulation codes are important in optimization of neutron optics of UCN sources and of experiments, but also in estimation of systematic effects, and in bench-marking of analysis codes. Here we will give a short overview of recent MC simulation activities in this field.
}

\kword{Monte Carlo simulations, ultracold neutrons, neutron electric dipole moment, neutron lifetime}

\begin{document}
\maketitle

\section{Introduction}

Since ultracold neutrons (UCN) have kinetic energies below 300\,neV, they experience total reflection on certain material walls (e.g. nickel) at any angle of incidence. The reflecting optical potential of wall materials originates from coherent nuclear scattering \cite{Fermi1936,Fermi1946} and must be larger than the kinetic energy of the UCN. This very low energy range is comparable to gravitational potential energy changes of the neutron. A UCN with 102.5\,neV can only reach a maximal height of one meter. A further interaction, the force exerted on UCN by a magnetic field gradient can also confine them in a volume for longer storage. A magnetic-field difference of 2 T along the UCN path causes a change of 120.6\,neV in its kinetic energy. The particle's magnetic moment couples to the magnetic field with a Larmor frequency of 29.2\,Hz in a field of 1\,$\mu$T allowing neutrons to be used as sensitive magnetometers. 

These interactions also allow UCN to be stored and manipulated, making them excellent probes in low-energy particle physics. The most important UCN experiments probe fundamental symmetries of nature (e.g. charge-parity violation by neutron electric dipole moment experiments \cite{Pendlebury2015,SerebrovEDM2015,Kasprzak2016}) or yield important high-precision parameters for Big Bang nucleosynthesis (e.g. neutron lifetime measurements \cite{Morris2017,Serebrov2008}). New, improved-precision experiments are in construction at all operating UCN sources around the world \cite{Bison2017}. 

Monte Carlo (MC) simulation methods are very effective in the characterization and optimization of many-parameter systems also in UCN-physics experiments. Several simulation codes are available e.g. STARucn \cite{Clement2014}, MCUCN \cite{Zsigmond2018}, PENTrack \cite{Schreyer2017}, and Geant4UCN \cite{G4UCN}. This allows for comparisons of intricate calculations, increasing confidence in these codes. 

The aim of this paper is to explain the most important features of STARucn, MCUCN and PENTrack, the codes which are extensively employed by our research groups and to briefly present inter-comparison examples. For Geant4UCN, used in one benchmark example, we refer the reader to \cite{G4UCN}. More details on the physics and the setup can be read in the references given for each example.

\section{Ultracold neutrons $-$ definitions used in simulations}
\label{sec-interaction}

Geometry implementations, trajectory calculations, spin precession calculations, and handling of input/output data are performed in various ways in these codes, as will be mentioned in the sections below. Interaction with magnetic fields will be described separately for each code.

The interaction of UCN with surfaces is treated per definition very similarly in all codes. This is based on the formalism of quantum reflection of waves on potential steps as used in cold neutron reflectometry \cite{Fermon1999}, however, also treating large incidence angles which are possible in the very low energy range of UCN  \cite{Golub1991}. 

Here we list the standard parameters related to surface interactions: 

\begin{itemize}
\item Volume-averaged, coherent nucleon-neutron-scattering potential (also known as Fermi or optical potential) 

\begin{equation}
    V_F = \frac{2 \pi \hbar^2}{m} N b_\text{coh} ,
\end{equation}

where $N$ is the density of nuclei, $b_\text{coh}$ the coherent-scattering length, and $m$ the mass of the neutron.
\item Material loss-constant $\eta = W_\text{loss} / V_\text{F} $, where 

\begin{equation}
W_{loss} =  \frac{\hbar}{2} N \sigma_{loss} v
\end{equation}

is the complex potential related to the absorption and up-scattering cross sections $\sigma_\text{loss}$, which is inversely proportional to the particle velocity $v$ in the medium. If the materials contain mixed scattering species $b_\text{coh}$ and $\sigma_\text{loss}$ must be replaced by their volume averages.  The energy- and direction-dependent neutron reflectivity is calculated from the complex modulus of the amplitude of the reflected wavefunction expressed in eq. (2.71) of \cite{Golub1991}

\begin{equation}
\label{eq-reflectivity}
		\left| R \right| ^2 = \frac
												{E_\perp - \sqrt{E_\perp} \sqrt{2\alpha-2\left(V_\text{F}-E_\perp \right)} + \alpha}
												{E_\perp + \sqrt{E_\perp} \sqrt{2\alpha-2\left(V_\text{F}-E_\perp \right)} + \alpha},
\end{equation} 

where $E_\perp$ is the perpendicular part of the kinetic energy, and 

\begin{equation}
\alpha = \sqrt{\left(V_\text{F}-E_\perp \right)^2+W_{loss}^2}. 
\end{equation} 

Eq. (\ref{eq-reflectivity}) is valid for all $E_\perp$ values both below and above the optical potential $V_\text{F}$.

\item For non-specular reflections a weighting parameter is most commonly used, which defines the fraction of ideal diffuse reflections (Lambert model) as discussed in subsection 4.4.5 in \cite{Golub1991}. Another approach is the so-called micro-roughness model formulated by \cite{Steyerl1972}, in which the interference of reflected waves is considered for roughness values comparable to the neutron wavelength.

\end{itemize}

\section{STARucn}
\label{sec-starucn}

The simulation package STARucn was originally developed at the Institut Laue Langevin for the purpose of simulating the GRANIT \cite{Roulier2015} experiment at the ILL. The original version is available and documented online \cite{Clement2014}, licensed under the open source GPLv3 license. It is heavily based on the ROOT package \cite{ROOT}, a modular scientific software framework developed for the experiments at CERN. Complex arbitrary geometries are supported. The software implements the propagation of UCN in gravity, interactions at material walls and volume interaction effects. STARucn can be programmed entirely with human-readable script files, and features an extensible plugin-based architecture, allowing additional features to be added with relative ease.

The package has since been forked and extended by a group at the University of Sussex, with the intention to apply it to simulations of UCN transport and systematic effects in the neutron electric dipole moment (nEDM) experiment at the Paul Scherrer Institut \cite{Kasprzak2016} and its successor, n2EDM. The largest change was the addition of a spin-tracking capacity, using a Runge-Kutta Cash-Karp 5th-order integrator with adaptive step size to track the precession of the neutron in magnetic fields, including relativistic effects from electric fields. The fields can be specified in a variety of ways: formula, map or several predefined common forms, any of which may be combined. A particular advantage for EDM experiments is that for each calculated trajectory, the spin may be tracked simultaneously in both directions of electric field, allowing simple calculations of E-field dependent frequency shifts. The software has been successfully used to model depolarization effects and relaxation times in nEDM experiments. Simulations of the geometric-phase effect \cite{Pendlebury2004} have been validated against MCUCN (see Sec.\,\ref{sec-intercomparison}) and analytical results.

Development efforts on this package focus on the inclusion of the effect of large magnetic-field gradients on the trajectory, the effect of finite mean free path due to a buffer gas, and extension of the simulation to support other species, for example mercury-199 (used as a comagnetometer in many nEDM experiments) or helium-3 (also popular as a magnetometer).

\section{MCUCN}
\label{sec-mcucn}

The MCUCN code was developed at PSI, supporting the projects of this group. It was extensively used in the optimization of the UCN source, beamlines and experiments at PSI to maximize UCN density and transmission \cite{Lauss2014}. MCUCN has already been a useful tool for estimating systematic effects in the neutron electric dipole moment (nEDM) experiment \cite{Pendlebury2015,Afach2015PRL,Afach2015PRD}, for checking data analysis software by toy data, or as part of the data analysis (providing e.g. collision rates \cite{Bondar2017}). Future MC simulations will also be deployed for helping the planning of time-efficient experiments e.g. by improving the online analysis with MC data. 

MCUCN is a pure C++ code and thus platform independent. Visualization of the MCUCN output data and geometry is left to the user who can choose her or his favorite tool. These features make MCUCN very flexible for calculations on computing grids. We have used extensively the PL-Grid \cite{PLGrid} in our applications. 

A storage volume or UCN-guiding geometry can be modeled using a combination of an arbitrary large number of reflecting or transmitting second-order surface sections (planes, cylinders, cones, etc.). Triangulated mesh exports from CAD programs in form of STL ASCII files \cite{STL} can also be used as input for MCUCN, however, an increase in computing time has to be taken into consideration.

Gravity playing an important role in UCN kinetics had to be implemented as well. In MCUCN, in contrast to Geant4UCN and PENTrack, the ballistic trajectories are analytically calculated between successive reflection or transmission points, making MCUCN calculations very fast. 

Motion in a strong magnetic field deflecting the UCN trajectories has not yet been generally implemented in MCUCN, but a local module can be called to calculate the UCN path in a cylindrical volume representing a strong local field like a magnetic valve. This module introduces small trajectory steps and thus strongly decreases  the computing speed. 

One important application of MCUCN was supporting nEDM-systematics calculations and related effects. Here we show an example for which the physics part related to gravitational resonance-frequency shift and depolarization was treated in detail in \cite{Pendlebury2015,Afach2015PRD}. We considered a cylindrical chamber ($H = $120\,mm, $r = $235\,mm) with its axis aligned with gravity, a geometry used in the nEDM experiment \cite{Pendlebury2015}. 

\begin{figure*}[htb]
\begin{center}
\resizebox{0.70\textwidth}{!}{
\includegraphics{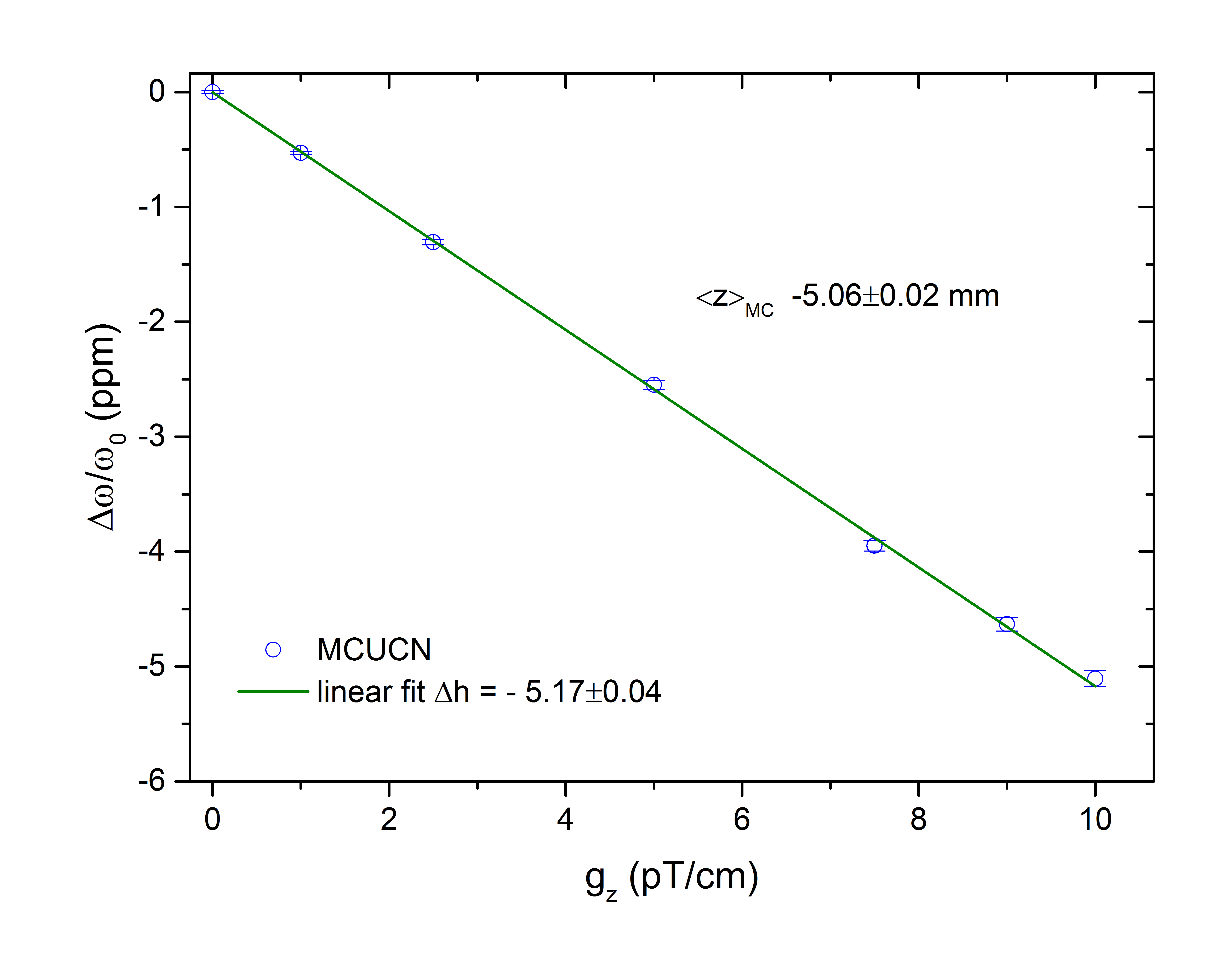}
}
\caption{Simulated relative frequency-shift (in ppm) due to gravitational depolarization as a function of the linear magnetic-field gradient $g_z$. The center-of-mass offset was obtained by a linear fit matching the simulated average vertical position within 2$\sigma$ (see text).} 
\label{fig:MCUCN-R-curve-trumpet-gz-fit}
\end{center}
\end{figure*}

As first described in \cite{Baker2006}, in the nEDM experiment the signal is interpolated to zero magnetic-field gradient. This can be done making use of the gravitational shift caused by a linear magnetic-field gradient parallel to gravity. The gradient must be modulated for both directions of the applied main magnetic-field vector (parallel and anti-parallel to gravity). A linear interpolation method can be used by the expression $\Delta \omega / \omega_0 = h/B_0\ g_z$, where $h$ denotes the center of mass offset from the cylinder center and $g_z=\partial _z B_z$ a constant gradient. In Fig.\,\ref{fig:MCUCN-R-curve-trumpet-gz-fit} we plotted the simulated frequency shift due the the gravitational effect and compare the extracted center-of-mass offset of UCN with the direct directly averaged value $\left\langle z \right\rangle _{MC}$ indicated above curve.  The two values match very well when fitting in the indicated range 0 - 10\,pT/cm, also attesting the reliability of the MCUCN algorithm.

In \cite{Harris2014} it was demonstrated theoretically that the linear approximation used earlier will not work for larger gradients above 25\,pT/cm, see also \cite{Zsigmond2018} for related simulations. 
 
\section{PENTrack}
\label{sec-pentrack}

PENTrack was initially developed at Technical University of Munich for the PENeLOPE experiment \cite{Materne2009} and is now also extensively used at TRIUMF for optimizations of a new UCN source and nEDM experiment \cite{Picker2017}. The implemented physics processes cover UCN transport, UCN storage in material bottles and magnetic traps, spin precession of neutrons and co-magnetometer atoms, and tracking of protons and electrons in electromagnetic fields. It provides a flexible configuration interface and allows to load complex electromagnetic fields and experiment geometries directly from finite-element (FEM) and computer-aided-design (CAD) software.

To track particles, PENTrack performs a 5th-order-Runge-Kutta integration of a general, relativistic equation of motion including gravitational acceleration, Lorentz forces due to magnetic and electric fields, and the force of a magnetic-gradient field on the particle's magnetic moment. To simulate the precession of spins in magnetic fields, PENTrack can integrate the Bargmann-Michel-Telegdi (BMT) equation along a particle's trajectory.

PENTrack allows to load maps of magnetic and electric fields calculated with commercial FEM tools on regularly spaced grids. Both three-dimensional maps of arbitrary fields and two-dimensional maps of rotationally symmetric fields are supported. Field values between grid points are calculated with tri- and bicubic interpolation  It also implements analytic fields, e.g. nearly homogeneous fields with small gradients and fields of straight conductors. Arbitrary time dependence of the fields can be described by a user-defined function.

Experiment geometries can be imported from virtually all CAD software as StL files. StL files approximate surfaces with triangle meshes. The intersection of a particle trajectory with such a surface is detected using a fast AABB-tree algorithm. Each part of the geometry can be deactivated in user-defined time intervals, making it possible to simulate variable properties of valves and other moving parts.

All simulation parameters---material properties, geometric model files, field maps, particle sources, and particle spectra---are stored in simple text files, allowing quick changes and optimization. PENTrack is written in C++ and its object-oriented structure simplifies the implementation of new electromagnetic fields, particles, and physics processes. It is based on open-source libraries and freely available at github.com/wschreyer/PENTrack.git

\begin{figure}[tbh]
\begin{center}
\resizebox{0.70\textwidth}{!}{	\includegraphics[width=\linewidth]{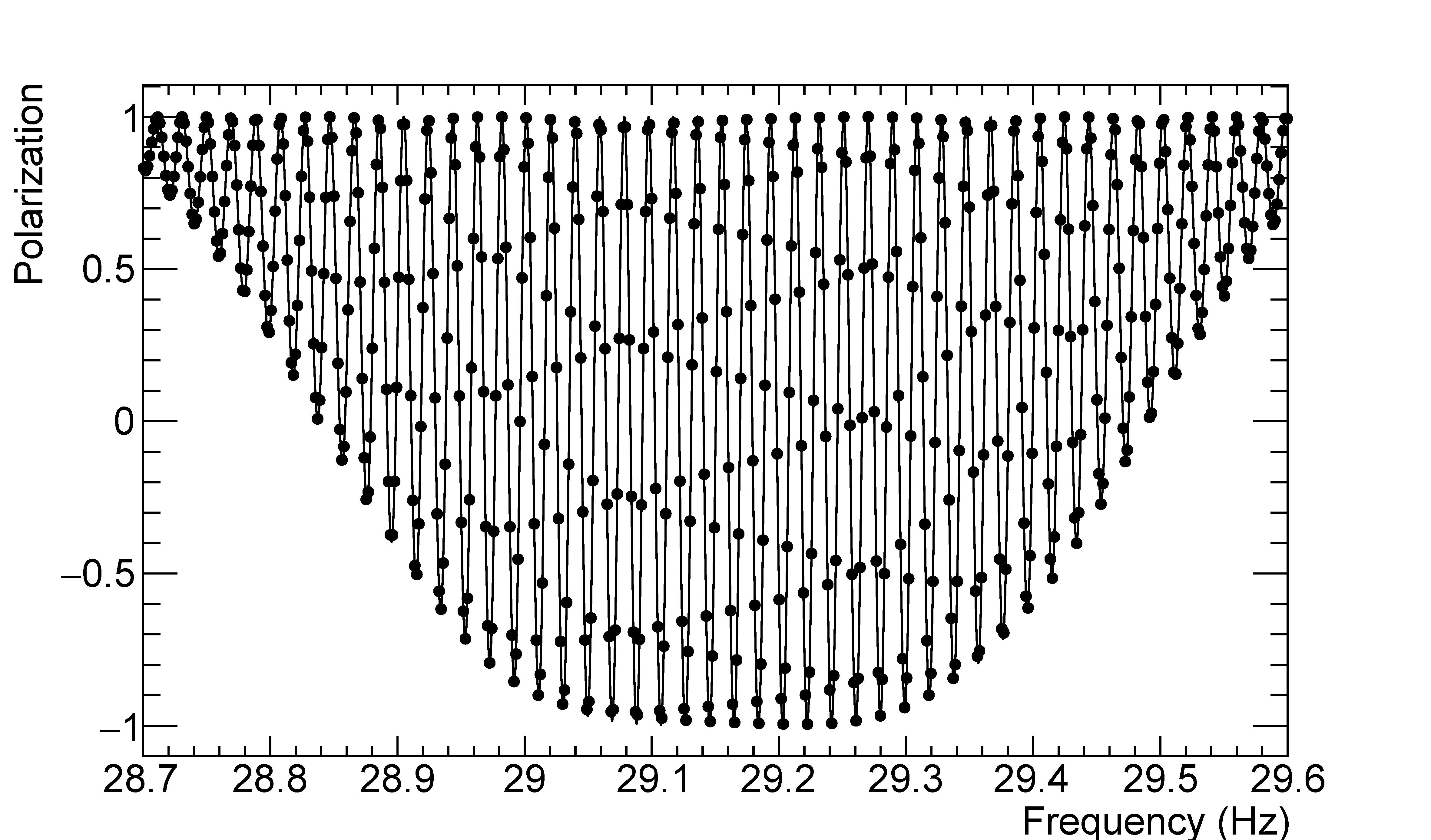}  }
	\caption{Polarization of a neutron spin after typical nEDM-experiment cycles with varying \(\pi/2\)-pulse frequency. The results simulated with PENTrack (dots) perfectly lie on the expected Ramsey pattern (solid line).}
	\label{fig:Ramsey_pattern}
\end{center}
\end{figure}

All aspects of PENTrack have been validated by comparing results with other simulation tools (see Sec.\,\ref{sec-intercomparison}), analytic calculations (see Fig.\,\ref{fig:Ramsey_pattern}), and experiments. In all cases the results showed very good agreement.

PENTrack is an excellent tool to study magnetic trap concepts for UCN like in the case of the PENeLOPE experiment \cite{Materne2009}. Protons and relativistic electrons from decays of magnetically stored UCNs, used to determine the neutron lifetime, can also be simulated. For nEDM experiments, the fully relativistic BMT equation allows simulation of geometric phases due to relativistic effects. For more details on PENTrack applications we refer the reader to \cite{Schreyer2017}.

\section{Code intercomparison examples}
\label{sec-intercomparison}

As mentioned in the introduction, it is very important to benchmark the simulation codes against each other. Benchmarks with analytic models and code intercomparisons increase confidence in the reliability of the codes that algorithms were implemented correctly or when intricate calculations have no other means for verification.

In the following, three simulation examples involving several codes are presented. For details and further discussion on the physics we refer the reader to the references given at each example.  

\begin{figure}[tbh]
\begin{center}
\resizebox{0.70\textwidth}{!}{	\includegraphics[width=\linewidth]{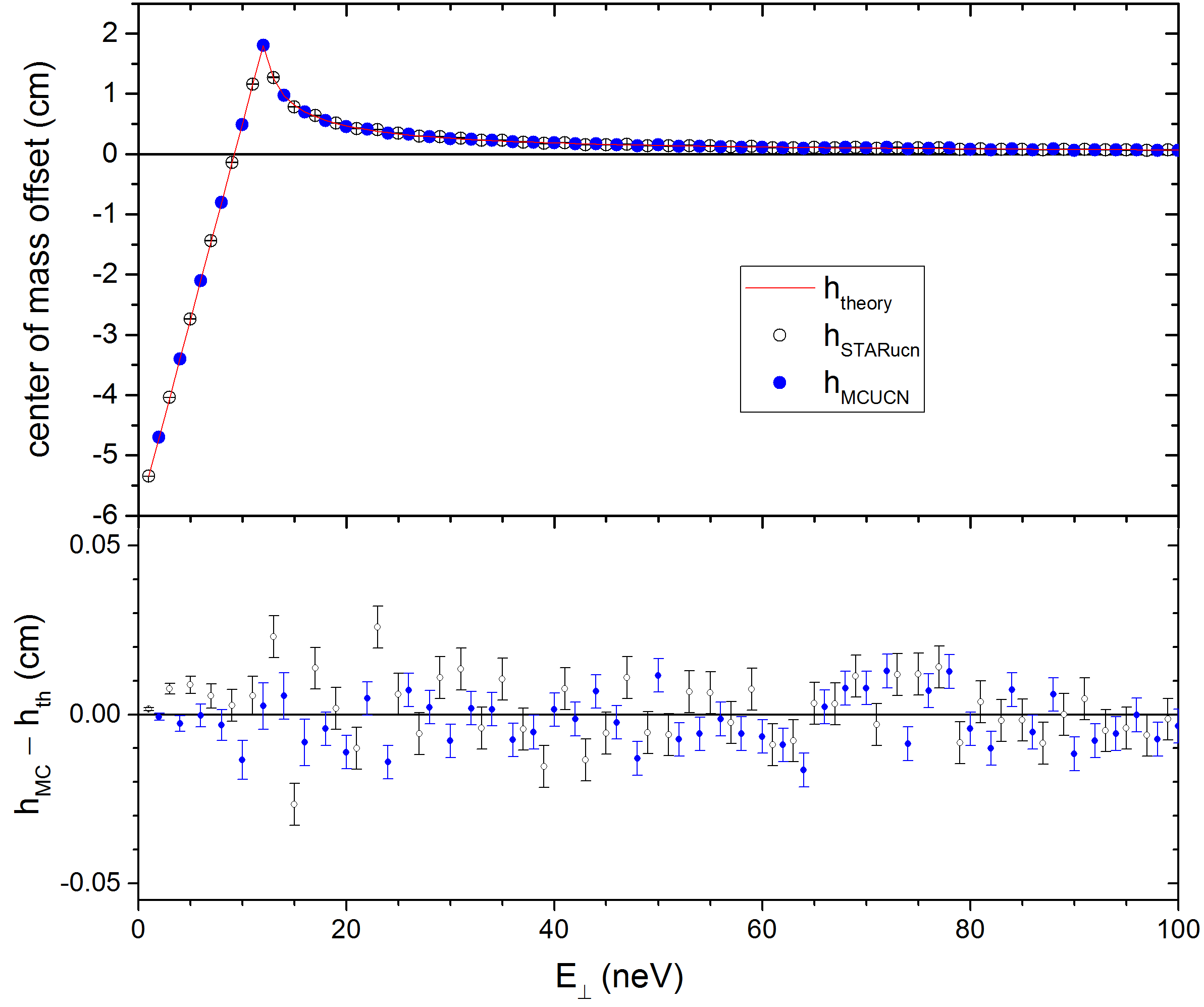}  }
	\caption{Comparison of MCUCN and STARucn simulations with analytic calculations of the energy-dependence of the center-of-mass offset for only specularly reflected UCN.}
	\label{fig:MCUCN-STARucn-Analytic-GravityNoDiffuse}
\end{center}
\end{figure}

As a check of the implementation of the equations of motion under influence of gravity in STARucn and MCUCN the average vertical coordinate of UCN i.e the center of mass offset in a cylindrical chamber (see also Sec.\,\ref{sec-mcucn}) was plotted in Fig. \ref{fig:MCUCN-STARucn-Analytic-GravityNoDiffuse} for the case of only specular reflections. For more details of the analytic model see \cite{Zsigmond2018}. Both codes reproduce very well the analytic calculations.

\begin{figure}[tbh]
\begin{center}
\resizebox{0.70\textwidth}{!}{  \includegraphics[width=\linewidth]{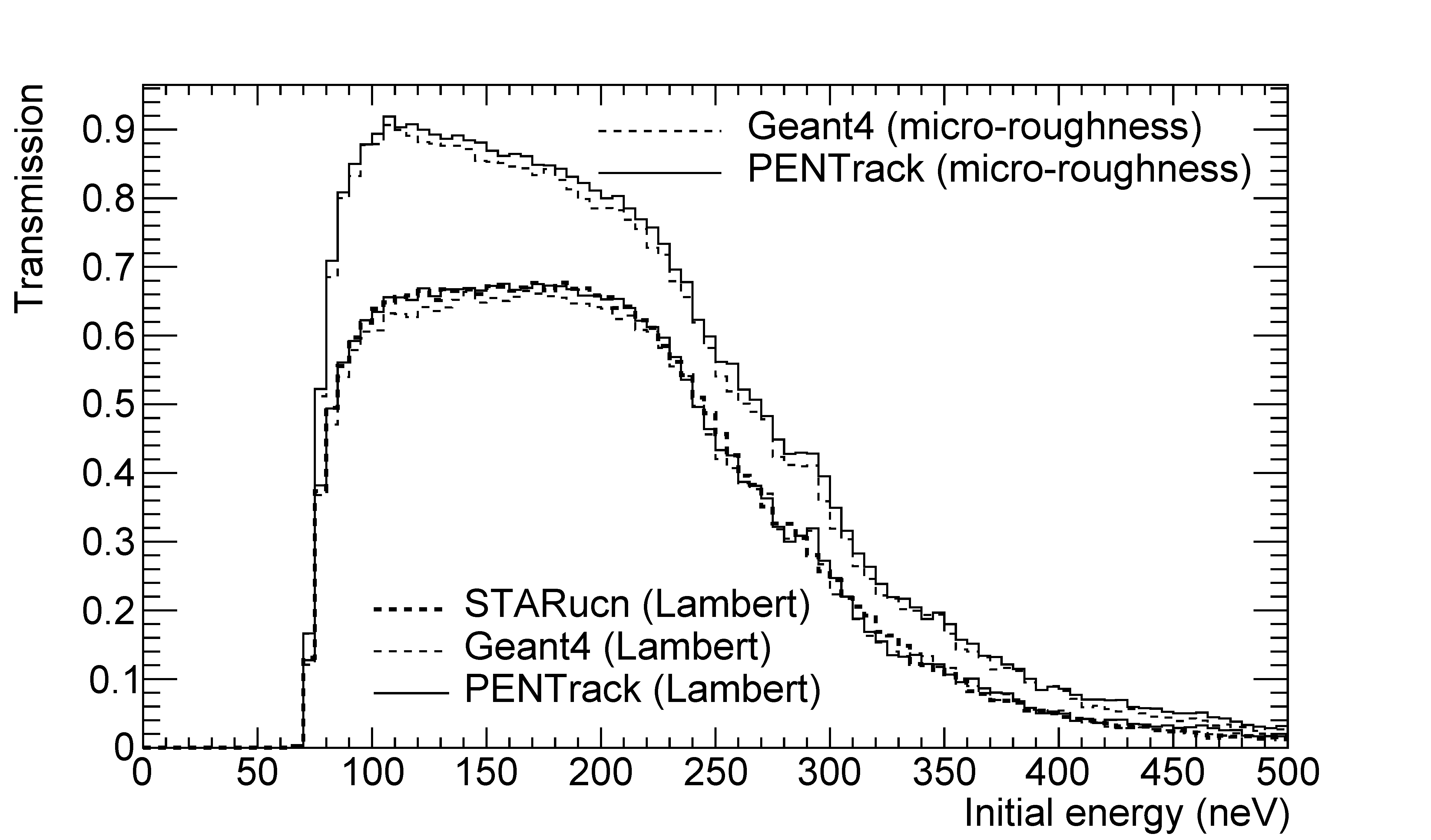}  }
\caption{Transmission of UCNs through a U-shaped guide, simulated with PENTrack, Geant4UCN, and STARucn. The UCNs have to overcome a vertical distance of 700~mm to traverse the guide, cutting off transmission below energies of 70~neV. STARucn does not support microroughness reflection.}
\label{fig:U-transm}
\end{center}
\end{figure}

In Fig. \ref{fig:U-transm} another intercomparison  is shown in which  the UCN transmission of a U-shaped guide was calculated as described in detail in \cite{Schreyer2017}. Using both Lambert and micro-roughness reflection models, PENTrack, STARucn, and Geant4UCN are in very good agreement. 

\begin{figure}[tbh]
\begin{center}
\resizebox{0.70\textwidth}{!}{  \includegraphics[width=\linewidth]{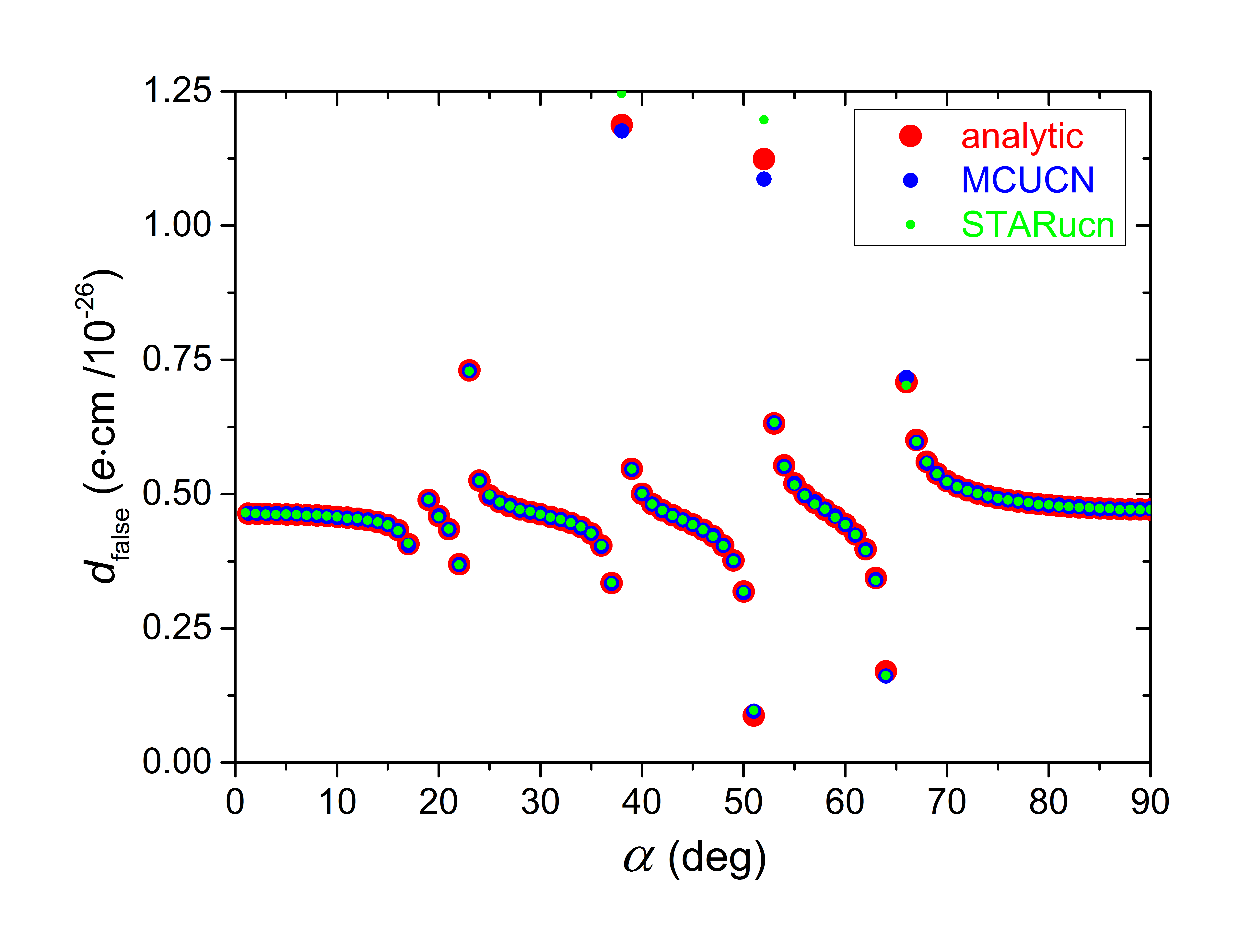}  }
\caption{Analytic calculations, STARucn simulations, and MCUCN simulations of the false nEDM signal as a function of the orbit section angle, as depicted in Fig. 3 of \cite{Pendlebury2004} (see text).}
\label{fig:MCUCN-STARucn-Eq-78-PendleburyPhysRevA70-2004-032102}
\end{center}
\end{figure}

In the last example, the simulated data points in Fig. \ref{fig:MCUCN-STARucn-Eq-78-PendleburyPhysRevA70-2004-032102} reproduce very well the analytic calculation according to eq. (78) in \cite{Pendlebury2004}. UCN orbiting with only specular reflections with a section alpha in a cylindrical precession chamber produce a false nEDM signal. This false nEDM signal originates from the interplay of transverse magnetic field gradients and a relativistic effect, the motional magnetic field $\textbf{B}_v = (\textbf{E}\times \textbf{v})/c^2$ seen by the UCN in the presence of an electric field (\textbf{E}), where v is the velocity vector of the UCN. The electric field is parallel or anti-parallel to the main $B_0$ field.

\section{Conclusions}

Since UCN experiments involve complicated particle trajectories (due to gravity and large solid angles), multiple surface interactions, and spin-precession in inhomogeneous magnetic fields, Monte Carlo simulations are an indispensable tool for UCN optics and data analysis optimizations and also directly provide input for data analysis. 

Parallel development of several UCN codes allows for important checks via code intercomparisons, but also motivates friendly competition supporting different approaches, and complementary ideas in treating various problems in UCN physics. 

\section*{Acknowledgements}

We gratefully acknowledge support from the UK Science and Technology Facilities Council, under Grants ST/ K001329/1, ST/M003426/1, and ST/L006472/1.  University of Sussex students Shane Black and Oliver Winston contributed to early development of the spin-tracking facility of STARucn. We also acknowledge granting access to the computing grid infrastructure PL-Grid\cite{PLGrid} used for MCUCN simulations. The development of PENTrack was supported by priority program SPP1491 ''Precision experiments in particle- and astrophysics with cold and ultracold neutrons'' of Deutsche Forschungsgemeinschaft and the Cluster of Excellence Exc153 ''Origin and Structure of the Universe''.


\begin{thebibliography}{9}
\bibitem{Fermi1936} E. Fermi,  Ric. Sci. \textbf{7}, 13 (1936)
\bibitem{Fermi1946} E. Fermi and W. H. Zinn, Manhattan District Declassified Code (1946)
\bibitem{Pendlebury2015} J. M. Pendlebury et al., Phys. Rev. D \textbf{92}, 092003 (2015).
\bibitem{SerebrovEDM2015} A. Serebrov et al., Phys. Rev. C \textbf{92}, 055501 (2015).
\bibitem{Kasprzak2016} M. Kasprzak, Hyperfine Inter. \textbf{237}, 142 (2016).
\bibitem{Morris2017} C. L. Morris, Rev. of Sci. Instr. \textbf{88}, 053508 (2017).
\bibitem{Serebrov2008} A. Serebrov et al., Phys. Rev. C \textbf{78}, 035505 (2008).
\bibitem{Bison2017} G. Bison et al., Phys. Rev. C \textbf{95}, 045503 (2017).
\bibitem{Clement2014}  {B. Cl{\'{e}}ment and D. Roulier}, STARucn, 2014 {https://sourceforge.net/projects/starucn/}.
\bibitem{Zsigmond2018} G. Zsigmond, Nucl. Instr. and Meth. A \textbf{881}, 16 (2018).
\bibitem{Schreyer2017} W. Schreyer et al., Nucl. Instr. and Meth. A \textbf{858}, 123 (2017).
\bibitem{G4UCN} F. Atchison et al., Nucl. Instr. and Meth. A \textbf{552}, 513 (2005).
\bibitem{Fermon1999} C. Fermon, F. Ott and A. Menelle,  Neutron Reflectometry. In: X-ray and Neutron Reflectivity: Principles and Applications, Springer (1999).
\bibitem{Golub1991}  R. Golub, D.J. Richardson and S.K.Lamoreaux, Ultra-Cold Neutrons, Adam Hilger (1991).
\bibitem{Steyerl1972} A. Steyerl, Z. Physik \textbf{254}, 169 (1972).
\bibitem{Roulier2015} D. Roulier, et al., Adv. in High En. Phys.,\textbf{2015}, 730437 (2015).
\bibitem{ROOT} R. Brun and F. Rademakers, Nucl. Inst. and Meth. A \textbf{389}, 81 (1997). See also http://root.cern.ch/.
\bibitem{Pendlebury2004} J. M. Pendlebury et al., Phys. Rev. A \textbf{70}, 032102 (2004).
\bibitem{Lauss2014} B. Lauss, Phys. Proc. \textbf{51}, 98 (2014).
\bibitem{Afach2015PRL} S. Afach et al., Phys. Rev. Lett \textbf{115}, 162502 (2015).
\bibitem{Afach2015PRD} S. Afach et al., Phys. Rev. D \textbf{92}, 052008 (2015).
\bibitem{Bondar2017} V. Bondar et al., Phys. Rev. C \textbf{96}, 035205 (2017).
\bibitem{PLGrid} Polish Grid Infrastructure, PL-Grid, http://www.plgrid.pl/en
\bibitem{STL} SLC File Specification, 3D Systems, Inc. (1994).
\bibitem{Baker2006} C. A. Baker et al., Phys. Rev. Lett \textbf{97}, 131801 (2006).
\bibitem{Harris2014} P. G. Harris et al., Phys. Rev. D \textbf{89}, 016011 (2014).
\bibitem{Picker2017} R. Picker, JPS Conf. Proc. \textbf{13}, 010005 (2017).
\bibitem{Agostinelli2003} S. Agostinelli et al., Nucl. Instr. and Meth. A \textbf{506}, 250 (2003).
\bibitem{Materne2009} S. Materne et al., Nucl. Instr. and Meth. A \textbf{611}, 176 (2009).
\end{thebibliography}
\end{document}